\documentclass[twocolumn]{aastex62}
\shorttitle{{\it SOFIA} FIFI-LS Observations of Sgr B1}
\shortauthors{Simpson et al.}

\usepackage{latexsym}
\usepackage{graphicx}
\usepackage{amssymb}
\usepackage{amsmath}
\usepackage{longtable}
\usepackage{epsf}

\begin{document}

\title{{\it SOFIA} FIFI-LS Observations of Sgr B1: Ionization Structure and Sources of Excitation
}

\author[0000-0001-8095-4610]{Janet P. Simpson}
\affiliation{SETI Institute \\
189 Bernardo Ave. \\
Mountain View, CA 94043, USA}
\email{janet.p.simpson@gmail.com}

\author{Sean W. J. Colgan}
\affiliation{NASA Ames Research Center \\
MS 245-6 \\
Moffett Field, CA 94035-1000}

\author{Angela S. Cotera}
\affiliation{SETI Institute \\
189 Bernardo Ave. \\
Mountain View, CA 94043, USA}

\author{Michael J. Kaufman}
\affiliation{San Jose State University \\
Department of Physics and Astronomy, One Washington Square \\ 
San Jose, CA 95192-0106}

\author{Susan R. Stolovy}
\affiliation{El Camino College \\
Physics Department, 16007 Crenshaw Blvd. \\ 
Torrance, CA 90506}

\begin{abstract}  % 250 words max

The current paradigm of Galactic Center (GC) gas motions and star formation 
envisions sequential star formation in streams of gas as they pass near the supermassive black hole, Sgr~A*.
This is based on the relative positions of dense molecular clouds, the very young star-forming region Sgr~B2, the much older region Sgr~C, and the several Myr old Arches and Quintuplet Clusters. 
Because Sgr~B1 is found with Sgr~B2 in a common envelope of molecular gas and far-infrared emission, the two sources are thought to be physically related, even though there are indicators of a significantly greater age for Sgr~B1. 
To clarify the status of Sgr~B1, we have mapped it with the FIFI-LS spectrometer on {\it SOFIA} in the far-infrared lines of [\ion{O}{3}] 52 and 88 \micron. 
From the ratios of these lines and lines measured with the {\it Spitzer} Infrared Spectrograph, 
we find that there are at least eight separate sub-regions that must contain the stars that excite the gas. 
We infer spectral energy distributions (SEDs) of the ionizing sources from models and find they 
are in agreement only with 
SEDs of late O stars augmented at the highest frequencies with interstellar X-rays from fast shocks. 
We suggest that although the gas, from its velocity structure, must be part of the very young Sgr~B2 complex, 
the stars that are ionizing the gas were not formed there but are the remnants of a previous generation of star formation in the GC.

\end{abstract}


\section{Introduction}

A recent model of star formation in the Galactic Center (GC) posits that 
there are streams of gas in open orbits around the nucleus (azimuthal period 3.69~Myr),  
with stars forming when the gas is compressed as it passes the pericenter, Sgr~A
(Longmore et al. 2013; Kruijssen et al. 2015).
In particular, for the dense molecular clouds at positive Galactic longitudes 
that can be described as lying on the front side of the orbit, the amount of star formation 
increases as a function of distance from Sgr~A (and hence time in the orbital model), 
with the most active star formation occurring in the very young Sgr~B2 region
(Longmore et al. 2013).
Farther along the orbit at negative Galactic longitude is found another massive 
but much older GC \ion{H}{2} region, Sgr~C. 
The Arches and Quintuplet Clusters formed from gas clouds that passed pericenter 
at even earlier times (Kruijssen et al. 2015). 

An anomaly in this scenario, however, is the luminous \ion{H}{2} region Sgr~B1, 
which appears to be part of the same giant molecular cloud as Sgr~B2 
but is already showing signs of dispersal as though due to winds from much earlier star formation
(e.g., Mehringer et al. 1992, hereafter M92). 
In the Kruijssen et al. (2015) model, Barnes et al. (2017) find that  
Sgr~B1 is on the back side of the orbit 
at an age of 1.5~Myr, versus 0.7~Myr for Sgr~B2. 

Even this age may be too young for Sgr~B1 --- 
recently Simpson (2018, hereafter S18) estimated an age of 4.6~Myr for Sgr~B1 
from her inferred shape of the spectral energy distribution (SED) of the stars 
that ionize this \ion{H}{2} region
(SEDs computed with Starburst99, Leitherer et al. 2014).
In S18, all the mid-infrared spectra of the GC taken by the Infrared Spectrograph with {\it Spitzer Space Telescope} were reanalyzed and 
the observed line ratios were compared to \ion{H}{2} region models computed with Cloudy 17.00 (Ferland et al. 2017). 
The results were that the Quintuplet Cluster region, the Arched Filaments, Sgr~B1, 
and Sgr~C \ion{H}{2} regions generally have the low excitation predicted by models 
ionized by SEDs of ages ranging from 2.5 to 5~Myr.
The data emphasized in the comparison were the [\ion{S}{3}] 33/[\ion{Si}{2}] 34~\micron\ line ratios, 
the [\ion{O}{4}] 26/[\ion{S}{3}] 33~\micron\ line ratios, 
and the [\ion{Ne}{3}] 15.6/[\ion{S}{3}] 18.7~\micron\ line ratios. 
%The first, a function of the ionization parameter (photon density$\div$electron density)  
The first, a function of the ionization parameter (photon density divided by the electron density)  
is an indicator of the dilution of the radiation field and hence shows the relative closeness 
of the ionizing stars. 
The second ratio, since O$^{3+}$ has an ionization potential ($IP$) of 54.9 eV, 
shows the high-energy content of the SEDs; 
S18 concluded that the SEDs ionizing GC \ion{H}{2} regions have X-rays additional to the 
Starburst99 high-energy photons. 
For the last, 
since the shape of the SED $> 13.6$~eV is a strong function of  
the stellar effective temperatures ($T_{\rm eff}$) (or age, for Starburst99 models),
ratios of the Ne$^{++}$ and S$^{++}$ ions, with $IP$ equal 41 and 23~eV respectively, 
indicate this shape and hence can be used to infer $T_{\rm eff}$ or age.

%and the relative numbers of ionizing photons in each energy interval can be estimated 
%from the intensities of forbidden lines that require particular photon energies 
%to ionize that energy level ($IP$), 
%the ionizing stellar $T_{\rm eff}$ or cluster age can be estimated 
%from ratios of lines with multiple $IP$.

In this Letter, we describe observations of Sgr~B1 produced in lines 
of O$^{++}$ (which has an intermediate $IP$ of 35 eV), 
made with the Field Imaging Far-Infrared Line Spectrometer 
(FIFI-LS; Colditz et al. 2018; Fischer et al. 2018) 
on the {\it Stratospheric Observatory for Infrared Astronomy} ({\it SOFIA}; Young et al. 2012; Temi et al. 2014).
Section~2 describes the observations, 
Section~3 depicts the computed electron densities ($N_e$), 
and the ratios of the ions O$^{++}$/S$^{++}$ and Ne$^{++}$/O$^{++}$, 
where the [\ion{Ne}{3}] and [\ion{S}{3}] line intensities were taken from S18.   
Section~4 compares the ionic abundance ratios and line ratios to models of \ion{H}{2} regions and shocks, 
and Section~5 presents the summary and conclusions. 
We will conclude that 
Sgr~B1 is ionized by widely-dispersed and relative cool OB stars with SEDs including X-rays,
possibly from high-velocity shocks. 

\section{Observations}

%Figure 1 ab
\begin{figure*}
%\includegraphics[width=184mm]{fig2_abc.eps}
%\plottwo{fig_radio_blue.eps}{fig_blue.eps}
\plottwo{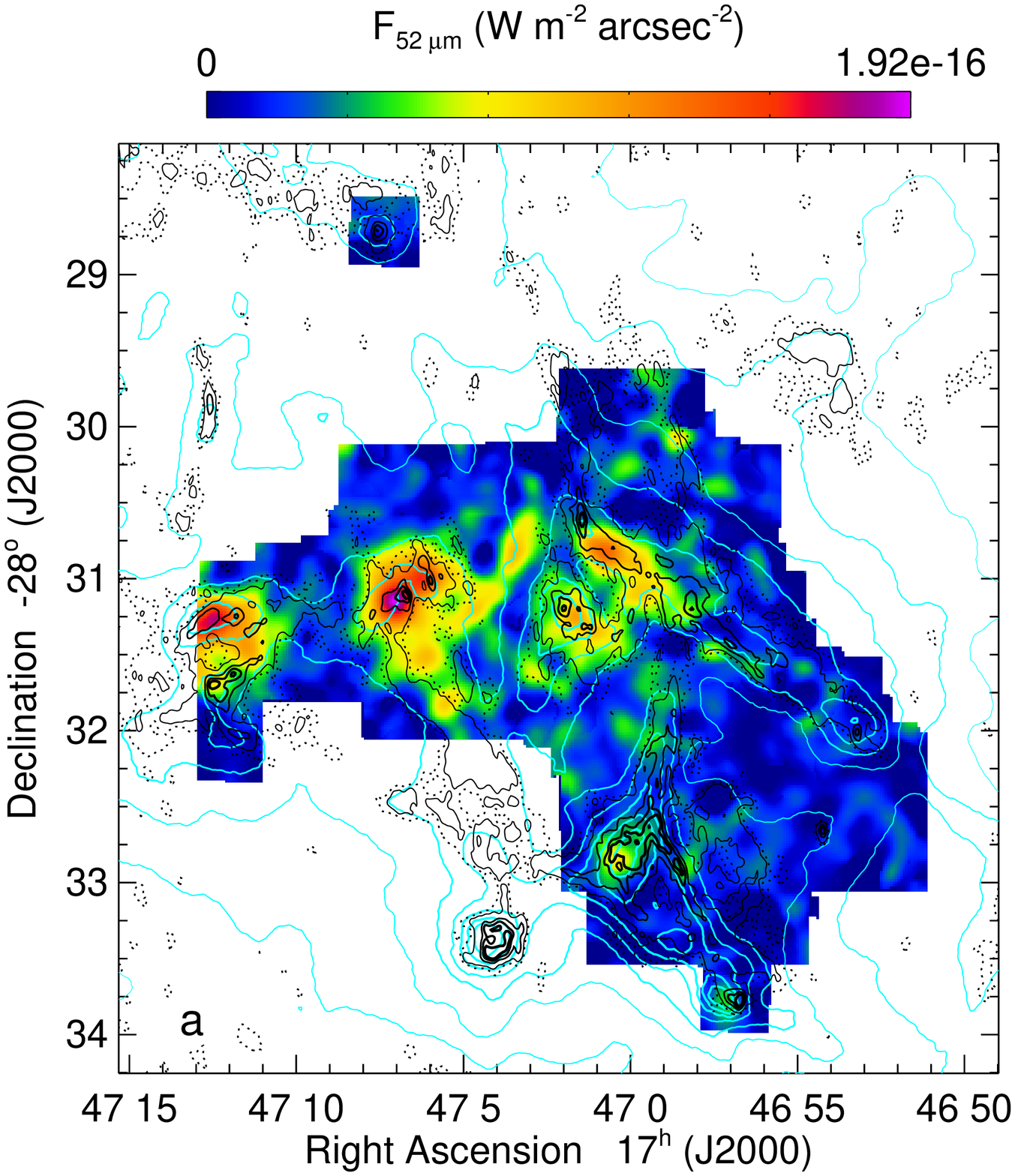}{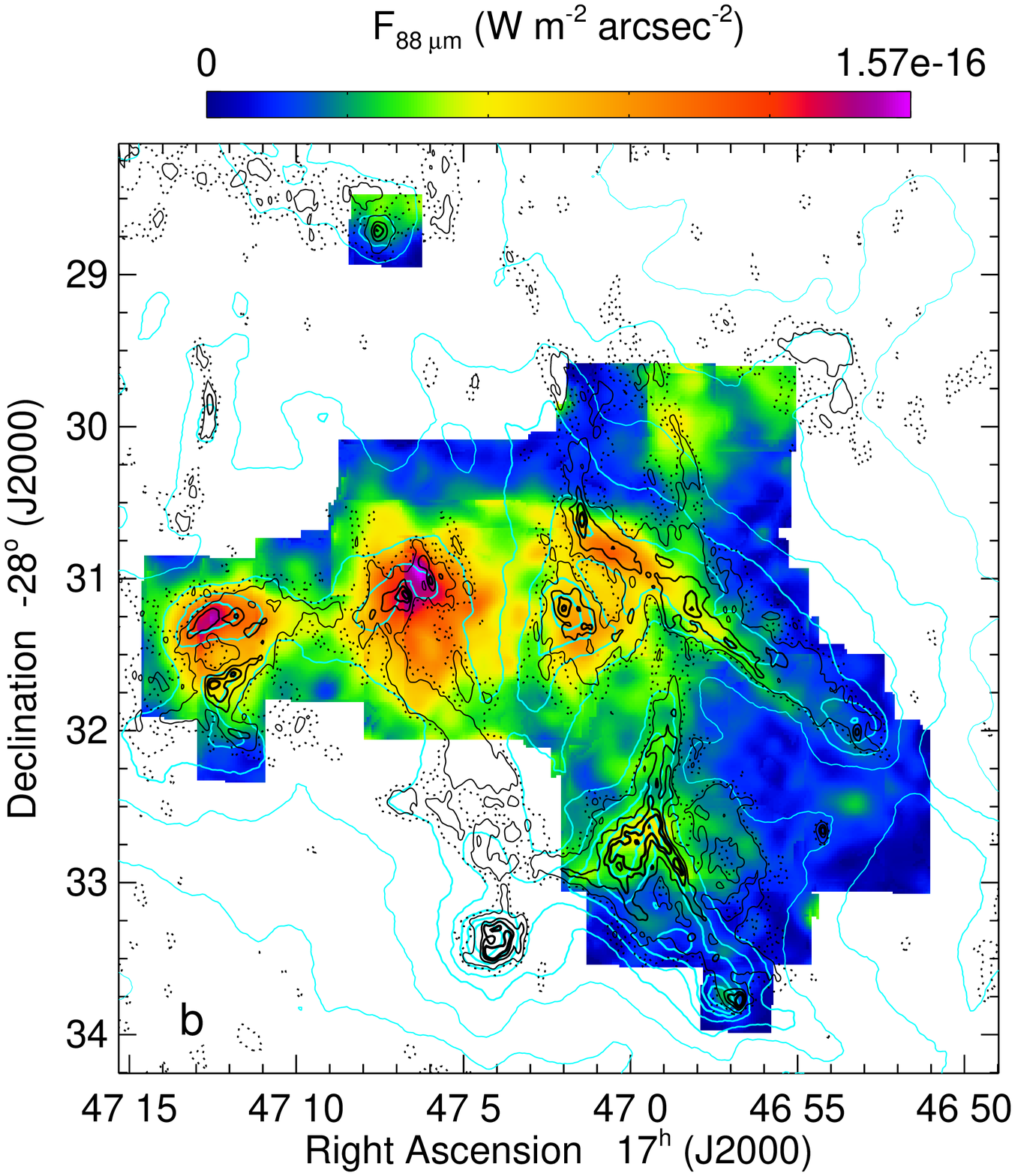}
\caption{
Observed [\ion{O}{3}] line intensities in Sgr~B1. All but the darkest blue pixels have S/N$ > 3$.
The black contours are the 8.4~GHz VLA intensities from M92 and the cyan contours are the 70~\micron\ {\it Herschel} Hi-GAL intensities (Molinari et al. 2011, 2016).
(a) The [\ion{O}{3}] 52 \micron\ line (maximum intensity $= 1.92 \times 10^{-16}$~W~m$^{-2}$~arcsec$^{-2}$).
(b) The [\ion{O}{3}] 88 \micron\ line (maximum intensity $= 1.57 \times 10^{-16}$~W~m$^{-2}$~arcsec$^{-2}$).
}
\end{figure*}

Sgr~B1 was mapped with FIFI-LS in July, 2016, and July, 2017, with {\it SOFIA} flying 
from Christchurch, NZ. 
FIFI-LS has two spectrometers, which operate simultaneously. 
The [\ion{O}{3}] lines at 51.81 and 88.36~\micron\ were observed with the blue channel (6\arcsec\ pixels, spectral resolution $\sim930$ and $\sim600$, respectively). 
The red channel lines ([\ion{O}{1}] 145.53~\micron\ and [\ion{C}{2}] 157.74~\micron) 
will be discussed in a later paper.

The parts of the map with the brightest radio emission were observed in 2016 and the rest in 2017.
%The parts of the map expected to be the brightest were observed in 2016; this map was 
%filled in and extended to the north in 2017 with some re-observation of the faintest sections.
In 2016, the chopper throw was 4\arcmin\ at position angles of 133 -- 155\degr\ 
east from north, 
approximately perpendicular to the Galactic plane.
In 2017, the chopper position angle for all observations was 135\degr\ 
with chopper throws 5\arcmin\ or 6\arcmin\ for the 52~\micron\ line 
and 6\farcm7 for the 88~\micron\ line.
All lines were measured in the `nod match chop' mode (telescope boresight 
half-way between the source and the reference positions); 
the resulting beam pattern is 
the diffraction-limited beam size (wavelength divided by 12, Young et al. 2012)
plus about 1\arcsec\ of coma added in quadrature 
for each arcmin of chopper throw.  

The FIFI-LS detectors form a grid of 5x5 pixels on the sky; 
sources are dithered with 3\arcsec\ steps and then 
the whole array is stepped by about 30\arcsec\ for another mini-map.
The FIFI-LS pipeline combines all the observations taken consecutively 
into a cube (RA, Dec, and wavelength) with 1\arcsec\ pixels.
To combine the resulting nine sub-cubes per line into a single cube, 
a large cube was first defined that covered the entire region as seen in the red channel in RA and Dec.
The wavelength dimension for this cube was that of the sub-cube with the maximum 
number of wavelength values (each of the sub-cubes had a slightly different wavelength scale).
Then the spectra of each of the sub-cubes were interpolated onto this wavelength array
(because the FIFI-LS line profiles are greatly oversampled, the interpolated fluxes 
were usually quite similar to the fluxes of their neighboring points).
Next, all the sub-cubes were shifted in RA and Dec onto the big grid. 
Because no interpolation was used, such shifts could be in error 
by as much as half a pixel (0\farcs5), small compared to the 7--10\arcsec\ spatial resolution.
Finally, all the new large cubes were combined --- where there were multiple integrations 
on a single point on the sky, the spectra were averaged and the errors were combined 
in quadrature. 
%(however, the pipeline errors are underestimated and will not be considered henceforth).

Line intensities were estimated for each spectrum, pixel by pixel on the sky,
by integrating the line profile over a given fixed range with a few pixels on each end defining the continuum.
Typically the number of wavelength pixels for the line was much larger than the number
for the continuum, especially for the 52~\micron\ line, which has a deep telluric H$_2$O 
line very close on the short wavelength side. 
The uncertainties for the line fluxes were estimated from the rms deviation of the data 
from a fitted line-profile function  
that gave a good representation of the overall shape of the spectrum but 
an unreliable flux. 
Because there are thousands of pixels, 
this fitting had to be done with an automated line-fitting program with little hand checking,
except for those measurements with signal/noise S/N$< 4$, which were checked by eye; 
very noisy line measurements were then rejected.

Maps of the 52~\micron\ and 88~\micron\ line intensities (linear scale) are plotted in Figure~1.
The measured continuum values are not useful because the small chopper throw in 2016 
resulted in the telescope chopping onto and subtracting
extra-source continuum (although not line, as estimated from the {\it Spitzer} maps of S18). 
In this figure, we see that the [\ion{O}{3}] morphology is distinctly different from both  
the 8.4~GHz radio map (proportional to the emission measure, $\int N_e^2 dl$) and
the 70~\micron\ {\it Herschel} image (proportional to the column density of warm dust),
which are quite similar. 

\section{Results}

%Figure 2 abcd
\begin{figure*}
%\includegraphics[width=184mm]{fig2_abcd.eps}
%\plottwo{fig_density.eps}{fig_SiS.eps}
%\plottwo{fig_OS.eps}{fig_NeO.eps}
\plottwo{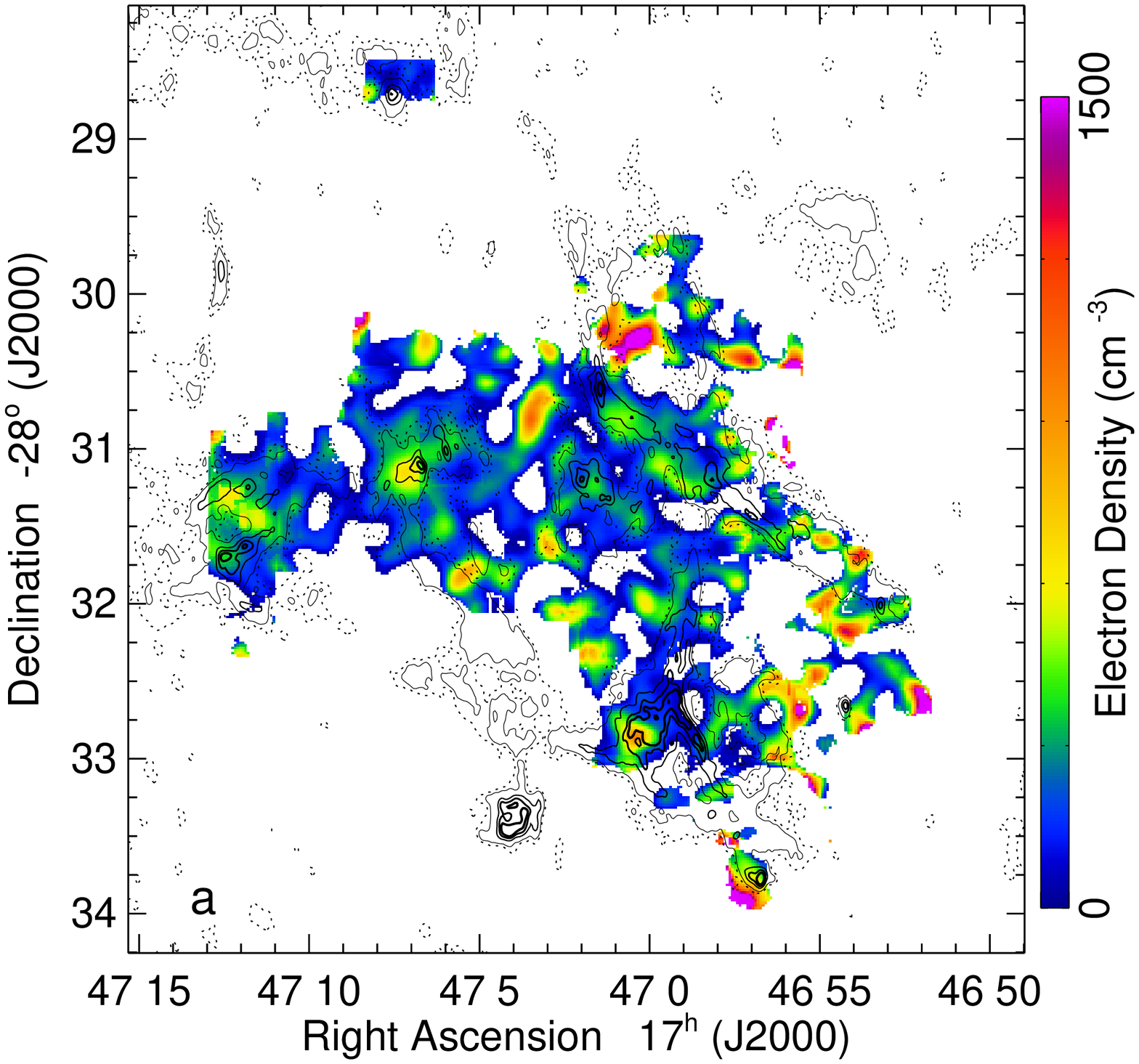}{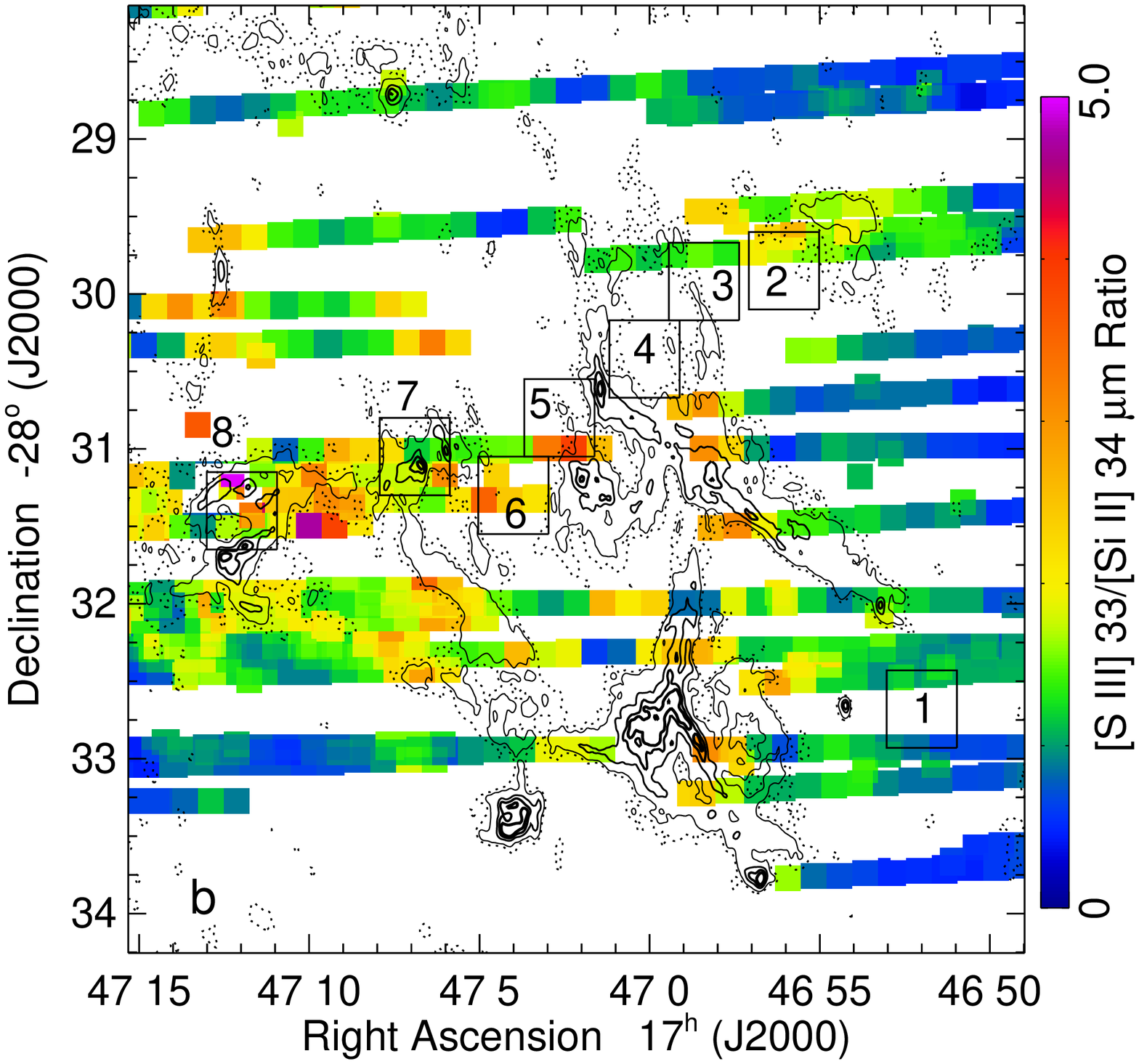}
\plottwo{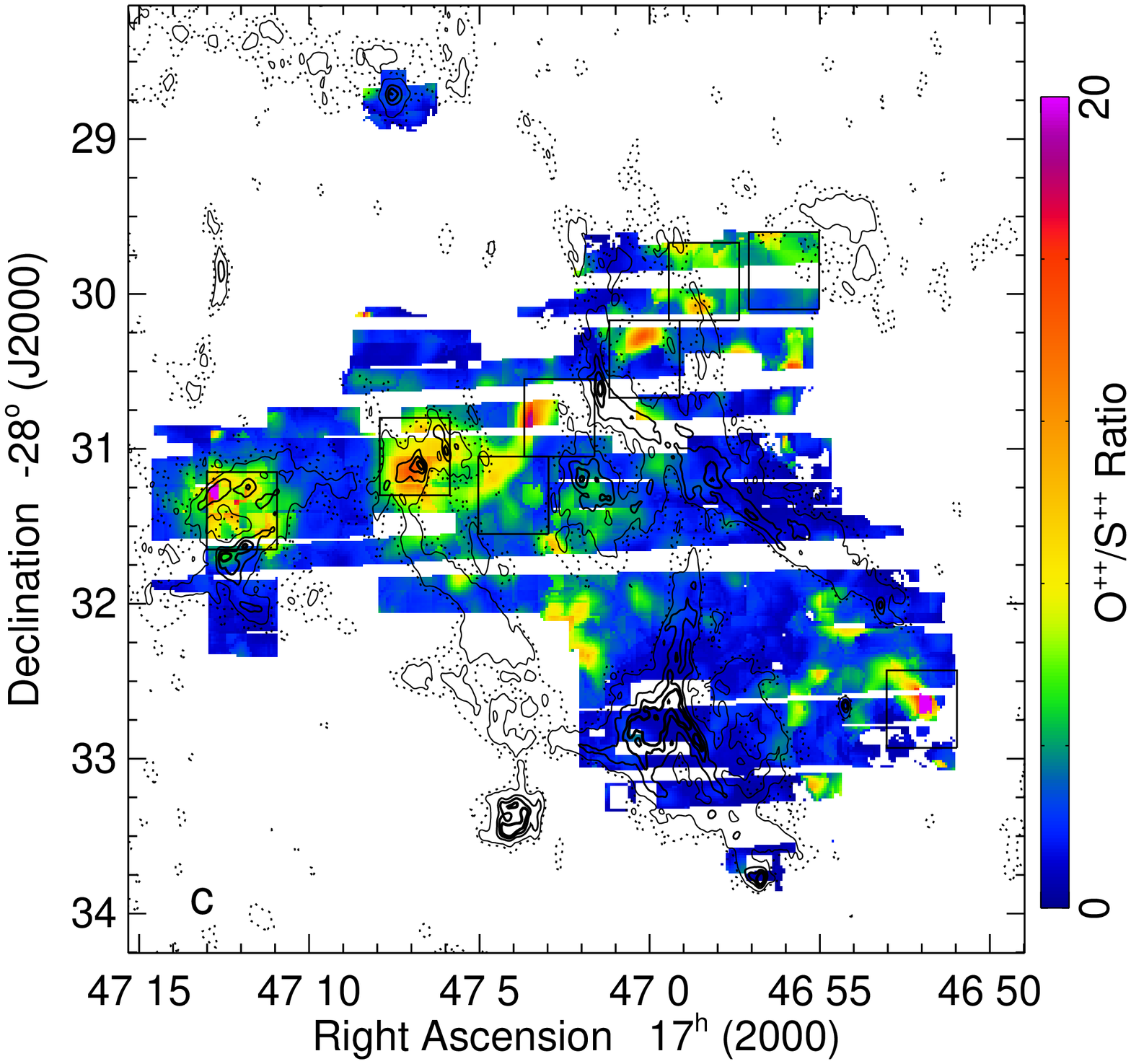}{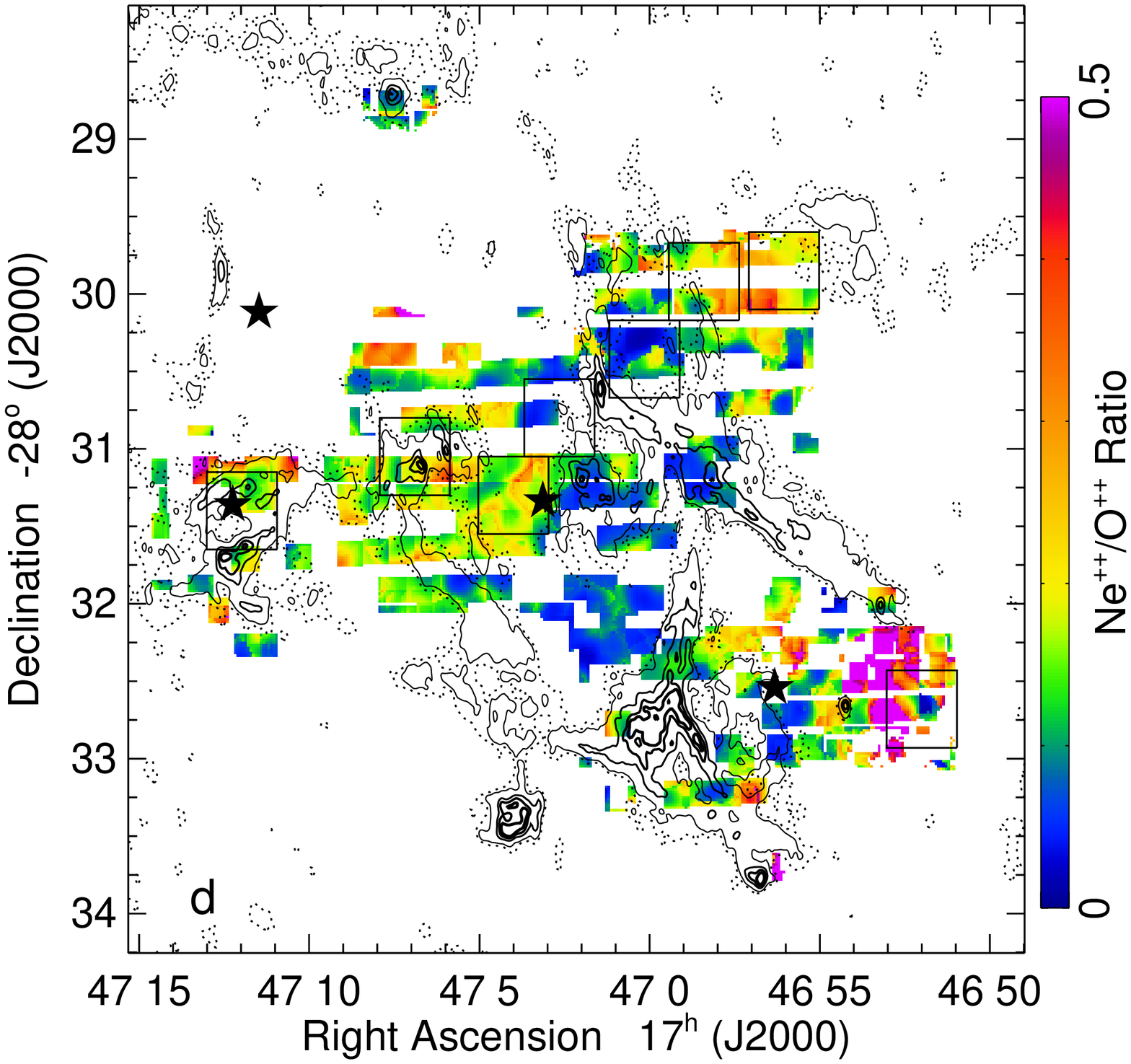}
\caption{Results from ionic abundance ratio calculations.
The contours in each map are those of the 8.4~GHz VLA image from M92.
(a) Computed electron densities from the [\ion{O}{3}] 52/[\ion{O}{3}] 88~\micron\ line ratios. 
The colored areas are where both [\ion{O}{3}] lines were observed  
and the line ratios have S/N~$ \geq 4$.
(b) The [\ion{S}{3}] 33/[\ion{Si}{2}] 34~\micron\ line ratio map from S18 (demonstrating the incomplete 
{\it Spitzer} map coverage).
The locations of the 8 sub-regions discussed in Section 4.1 are plotted as the numbered black boxes.
(c) The ionic O$^{++}$/S$^{++}$ ratio. 
(d) The ionic Ne$^{++}$/O$^{++}$ ratio. 
The large black stars show the locations of the O supergiant and the WR stars identified by Mauerhan et al. (2010).
}
\end{figure*}

We corrected the observed line intensities for the extinction estimated by S18
and computed ionic abundance ratios using the method described by Simpson et al. (2004, 2007, 2012), 
with O$^{++}$ collision strengths from Storey \& Sochi (2015)
and other atomic physics from Table~1 of S18.
For estimates of $N_e$, we required that the 52/88~\micron\ line ratio  
was measured with S/N$ \geq 4$ and
required line measurements with S/N$ \geq 3$ for ionic abundance ratios. 
Maps of these results (all linear scale) are plotted in Figure~2.

Figure~2(a) shows a plot of $N_e$ with the radio contours overplotted. 
{ 
The average density for the high S/N region is low, $N_e \sim 300$ cm$^{-3}$.
The regions with the highest intensity radio contours on the western side of Sgr~B1 --- the `Ionized Bar' (17$^{\rm h}$47$^{\rm m}$01\fs4~$-$28\degr~30\arcmin~37\arcsec) and the `Ionized Rim' (17$^{\rm h}$46$^{\rm m}$59\fs9~$-$28\degr~32\arcmin~53\arcsec) of M92 ---
are examples of low densities. 
Here, a substantial distance of ionized gas along the line of sight is required to produced the observed emission measure, given the low density. 
Low densities for this region are also estimated from the ratios of the lower-excitation [\ion{S}{3}] 19/[\ion{S}{3}] 33~\micron\ lines, where S$^{++}$ is the dominant ionization state of sulfur (S18). 
We speculate that these structures are edge-on sheets of gas surrounding local holes or bubbles blown by winds from hot stars. 
The eastern parts of Sgr~B1, however, have densities commensurate with the intensity of the radio emission.
The highest densities are found in regions of low intensity and may be external to Sgr~B1.
}

%Figure~2(a) shows a plot of $N_e$ with the radio contours overplotted. 
%We see that the highest intensity radio contours correspond to regions of low $N_e$, 
%from which we infer that the ionized gas is found in 
%layers with depth greater than width on the sky.
%layers of gas seen edge-on.
%The highest densities are found in regions of low intensity and may be external to Sgr~B1.
%Only the eastern parts have densities commensurate with the intensity of the radio emission.
%The average density for the high S/N region is $N_e \sim 300$ cm$^{-3}$.

%It is clear that $N_e$ does not track the emission measure --- 
%in fact, the density in the radio features 
%appears to be fairly constant (average value of  $N_e \sim 300$ cm$^{-3}$). 
%We infer that the regions of apparent increase in radio brightness are sheets of gas seen edge-on.

Figure~2(b) shows the [\ion{S}{3}] 33/[\ion{Si}{2}] 34~\micron\ line ratios from {\it Spitzer} (S18).
This ratio is a function of the photon density of the radiation field, and hence demonstrates that 
the exciting stars are widespread, with occasional large gaps (ratio $\lesssim 1.5$)
where there are few ionizing stars.

Figures~2(c) and 2(d) show the O$^{++}$/S$^{++}$ and Ne$^{++}$/O$^{++}$ ratios, respectively
(the O$^{++}$/S$^{++}$ ratio has better coverage than the Ne$^{++}$/O$^{++}$ ratio because we could use both the [\ion{S}{3}] 19 and 33~\micron\ lines).
These ratios are indicative of the shapes of the ionizing SEDs and so 
are an indication of the locations of the exciting stars with the highest $T_{\rm eff}$. 

\section{Discussion}

\setcounter{table}{0}
\begin{table*}
\centering
\begin{minipage}{160mm}
\caption{Positions Inferred to have Nearby Sources of Ionization
}
\begin{tabular}{@{}lcccccccc@{}}
\hline
\hline
Position & RA & Dec & log $N_{\rm Lyc}$\tablenotemark{a} & $N_e$ & O$^{++}$/S$^{++}$ & Ne$^{++}$/O$^{++}$ & $\frac{[{\rm O\ III}] 88~\micron}{[{\rm S\ III}] 33~\micron}$ & $\frac{[{\rm Ne\ III}]15.6~\micron}{[{\rm O\ III}] 88~\micron}$    \\
     & (J2000)  & (J2000)  & (s$^{-1}$) & (cm$^{-3}$) & & & line ratio & line ratio  \\
\hline
1 & 17 46 52.00 & $-$28 32 41 & 48.09 & 841 & 6.4 & 0.50 & 0.16 & 0.89  \\
2 & 17 46 56.05 & $-$28 29 51 & 48.27 & ... & 4.1 & 0.24 & 0.31 & 0.28  \\
3 & 17 46 58.40 & $-$28 29 55 & 48.29 & 269 & 5.7 & 0.21 & 0.38 & 0.30 \\
4 & 17 47 00.15 & $-$28 30.25 & 48.72 & 723 & 4.4 & 0.08 & 0.06 & 0.18  \\
5 & 17 47 02.65 & $-$28 30 48 & 48.70 & 297 & 7.5 & 0.09 & 0.17 & 0.20  \\
6 & 17 47 04.00 & $-$28 31 18 & 48.58 & 181 & 4.9 & 0.18 & 0.30 & 0.26 \\
7 & 17 47 06.90 & $-$28 31 03 & 48.73 & 342 & 8.7 & 0.19 & 0.34 & 0.42 \\
8 & 17 47 11.98 & $-$28 31 25 & 48.62 & 352 & 6.6 & 0.17 & 0.38 & 0.32 \\
\hline
\end{tabular}\
\tablecomments{
\tablenotetext{a}{The estimated numbers of photons required to ionize the sub-regions were derived using Equation (4) of Simpson et al. (2012), where the measured [\ion{S}{3}] 33~\micron\ line flux was integrated over the sub-region, the assumed S/H ratio was $1.90 \times 10^{-5}$, and the assumed S$^{++}$/S ratio was 0.8 (a likely value for these low-excitation \ion{H}{2} regions, which have much lower and more uncertain O$^{++}$/O).
The total estimated $N_{\rm Lyc}$ from [\ion{S}{3}] is $3.3 \times 10^{50}$ s$^{-1}$, in agreement with 
the estimated number of ionizing photons from the radio, $3 \times 10^{50}$ s$^{-1}$ (S18), for assumed distance 8~kpc.} 
}
\end{minipage}
\end{table*}

\subsection{Sources of Ionization}

We immediately notice that Sgr~B1 is not ionized by a central star cluster --- 
there are multiple regions of higher ionization, as seen in 
the O$^{++}$/S$^{++}$ and Ne$^{++}$/O$^{++}$ ratios plotted in Figures~2(c) and 2(d).
After smoothing the O$^{++}$/S$^{++}$ ratio by 30\arcsec, we selected the eight 
regions with highest O$^{++}$/S$^{++}$; 
they are delineated in Figure 2(b) and listed in Table~1.
There are undoubtedly additional sub-regions but the combined {\it Spitzer} and {\it SOFIA} data 
do not have enough coverage (e.g., Figure~2) to adequately define them. 

It is particularly interesting that these regions of higher ionization do not coincide 
with the peaks in the radio emission (contours in Figures~1 and 2).
M92 measured the flux density in each of these peaks and 
estimated the ZAMS spectral types of the OB stars that would be required to ionize them.
Our observations, which mostly locate the ionizing stars in volumes separate from the ionized gas, 
show that the ionizing stars must be positioned in some pattern other than at the peaks of the 
radio emission. 
For example, 
{ the `Ionized Rim' requires the most ionizing photons; it}
%For example, the region requiring the most ionizing photons, 
%the `Ionized Rim' at 17$^{\rm h}$46$^{\rm m}$59\fs9~$-$28\degr~32\arcmin~53\arcsec,
 certainly does not have the excitation corresponding to M92's suggested O6 star. 
Multiple stars of cooler temperature would be needed to ionize this region. 

The morphology of the extended features and shell structures in the radio emission 
led M92 to suggest that Sgr~B1 is an evolved \ion{H}{2} region. 
There certainly are no dense, ultracompact \ion{H}{2} regions like there are in Sgr~B2 
(e.g., Mehringer et al., 1993; De Pree et al. 2015), 
which is probably physically associated owing to the similar velocities 
and apparent gas structures that bridge the two regions (M92).

If Sgr~B1 is indeed an evolved \ion{H}{2} region, it should contain stars that have evolved significantly.
% and not ZAMS stars ionizing ultracompact \ion{H}{2} regions. 
Two to three Wolf-Rayet (WR) stars and an O supergiant were found in Sgr~B1 by Mauerhan et al. (2010) 
and are plotted in Figure~2(d).
%Only one of these, CXOGC J174712.2-283121, a WN7-8h star, is found in one of our denser sub-regions 
%of higher ionization with the others being found in low density, low emission regions. 
%These stars in the line of sight to Sgr~B1 might not be significant --- 
The presence of evolved stars in the line of sight to Sgr~B1 might not be significant --- 
Habibi et al. (2014) simulated orbits for stars drifting away from the Quintuplet and Arches Clusters
(ages $4.8 \pm 1.1$ and $3.5 \pm 0.7$ Myr, Schneider et al. 2014). 
They found that some of their simulated stars could travel as far from the Quintuplet Cluster as Sgr~B1. 

Another possibility could be that while the ionized and molecular gas is part of the Sgr~B 
molecular cloud (M92; Mehringer et al. 1995; Lang et al. 2010),
the ionizing stars originate in a much older cluster that has already orbited once around Sgr~A,
much as is thought to have occurred for the Quintuplet and Arches Clusters
(e.g., Habibi et al. 2014; Kruijssen et al. 2015).
These stars light up the edges of local molecular clouds (Lang et al. 2002) but are not near the locations of their formation (Stolte et al. 2014). 
We note that the multiple young stellar objects found in Sgr~B1 (An et al. 2011, 2017), none especially luminous,  
are all found in the denser of the radio peaks; 
we suggest these may be indicators of star formation triggered by the outflows from the very luminous ionizing stars.

\subsection{Ionizing SEDs}

%Figure 3 ab
\begin{figure*}
\centering
\includegraphics[width=125mm]{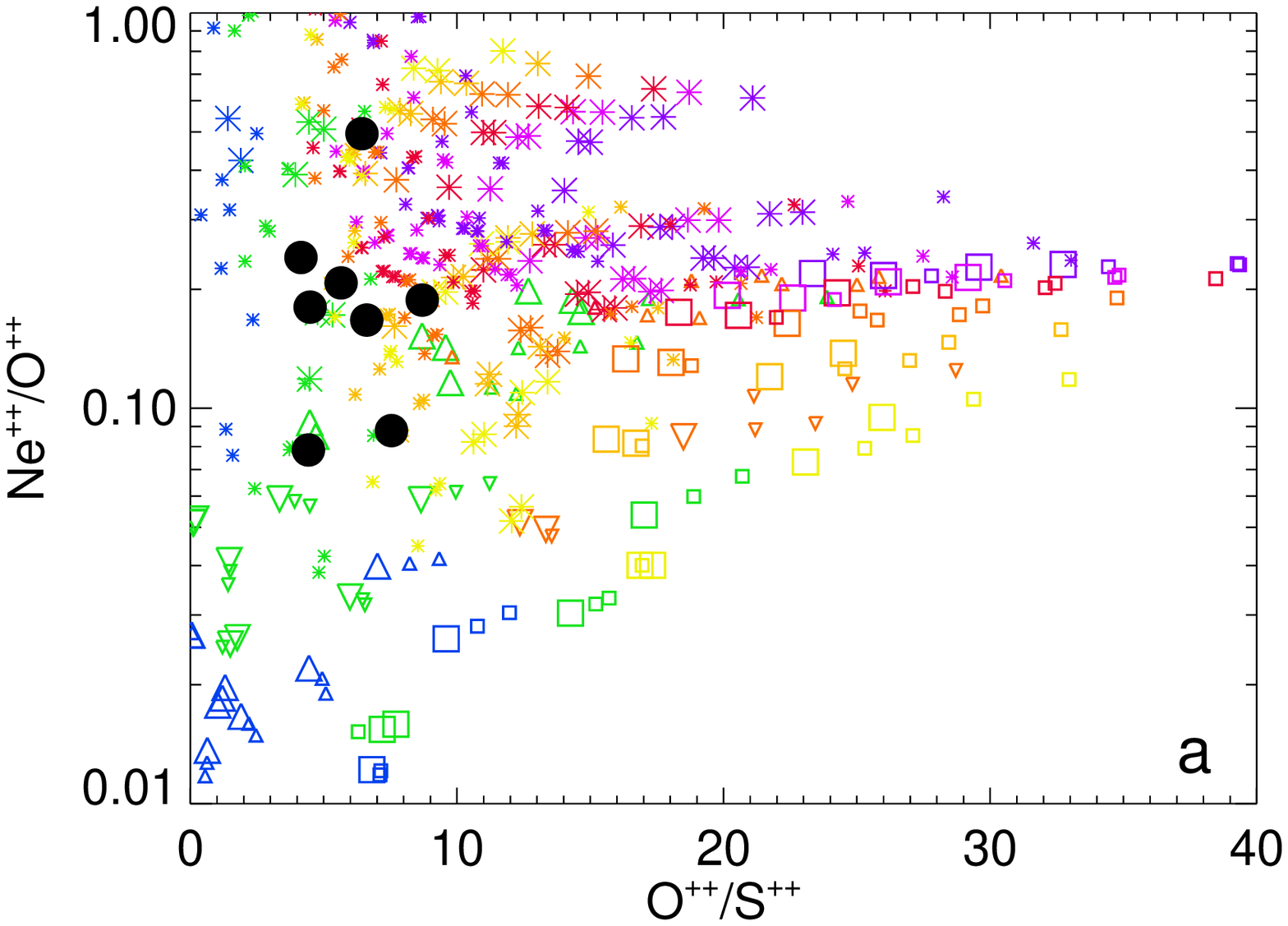}
\includegraphics[width=125mm]{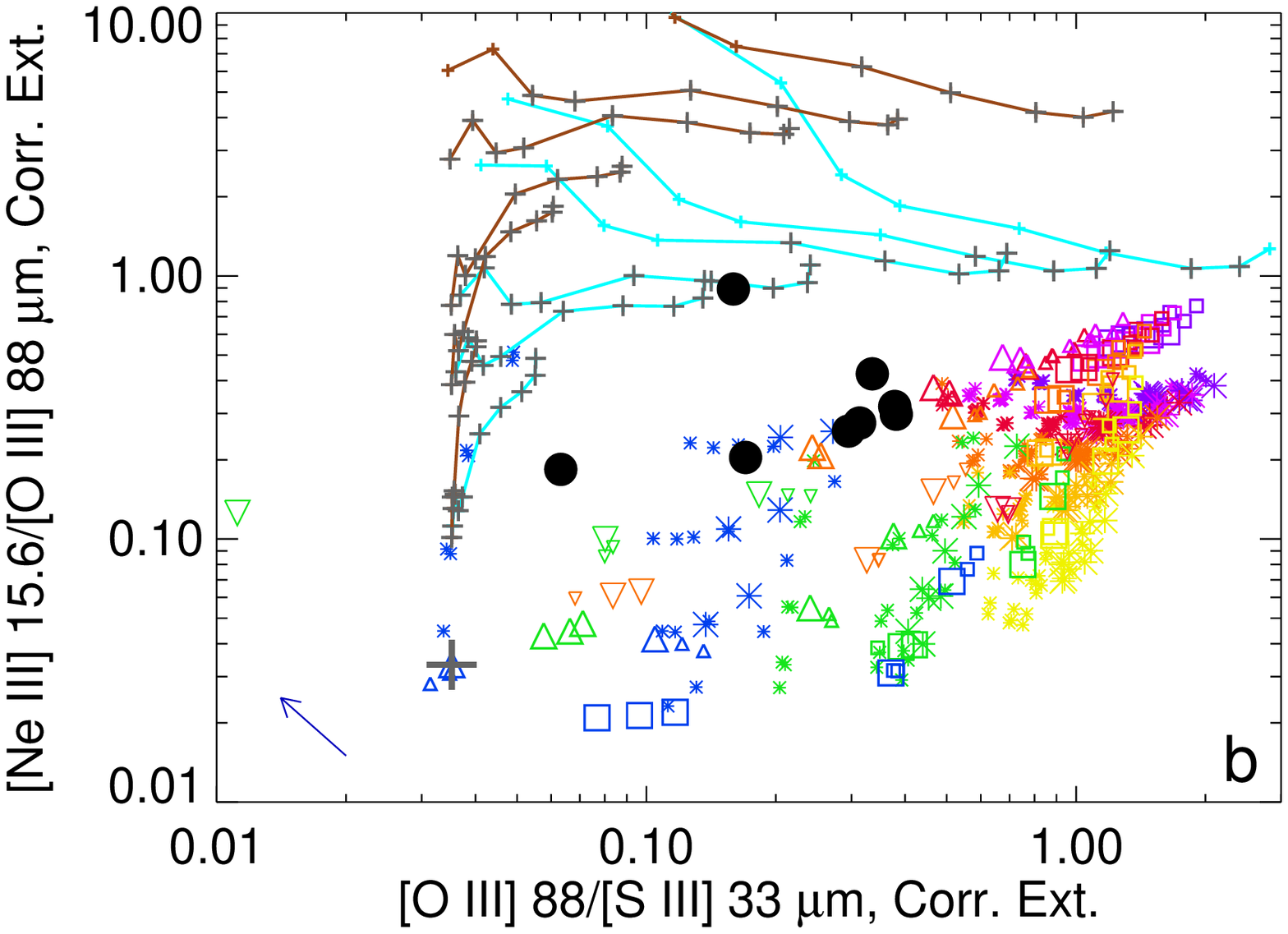}
%\plottwo{fig3a.eps}{fig3b.eps}
\caption{Ionic and line ratios compared to models.
In both panels, the black dots are our observed positions as described in the text and listed in Table~1.
The asterisks are models from S18 
computed with Cloudy (Ferland et al. 2017) with 
O/S=36 and Ne/O=0.25, 
ionizing SEDs from Starburst99 (Leitherer et al. 2014), and extra X-rays represented by 
blackbodies with $T_{\rm BB} = 10^{6.5}$~K, where the colors signify the ages of the Starburst99 
model SEDs: blue is $10^{6.7}$~yr, green is $10^{6.65}$~yr, yellow is $10^{6.6}$~yr,
yellow-orange is $10^{6.5}$~yr, red-orange is $10^{6.4}$~yr, red is $10^{6.3}$~yr, 
magenta is $10^{6.2}$~yr, and purple is $10^{6.0}$~yr.
Squares are newly computed models with input density $= 300$~cm$^{-3}$, the same Starburst99 SEDs 
and otherwise the same input parameters 
but no X-rays, and the same color scheme.
Upward- and downward-pointed triangles are similar models with supergiant and dwarf SEDs 
computed with WM-BASIC (Pauldrach et al. 2001) 
taken from { BPASS} (Eldridge et al. 2017); 
the colors represent $T_{\rm eff}$ equal to blue: 30200~K, green: 32300~K, orange: 34600~K, 
red-orange: 37200~K, and red: 40000~K.
For all models, the larger and smaller symbols represent models whose predicted [\ion{S}{3}] 33/[\ion{Si}{2}] 34~\micron\ line ratios are within or outside the range of the observed ratio, 1.5 to 3.0,
respectively.
(a) The Ne$^{++}$/O$^{++}$ ratio versus the O$^{++}$/S$^{++}$ ratio.
(b) The [\ion{Ne}{3}] 15.6/[\ion{O}{3}] 88~\micron\ line ratio versus the [\ion{O}{3}] 88/[\ion{S}{3}] 33~\micron\ line ratio,
where the line intensities were corrected for extinction from S18.
{ Recomputing the models plotted with asterisks for a higher $N_p = 300$~cm$^{-3}$ would have them move in the direction of the blue arrow.}
Following Ho et al. (2014), the { cyan and brown} lines are the result of adding shock emission from Allen et al. (2008) { (see text)}. 
{ The gray crosses are combinations of models plus shocks} that predict the [\ion{S}{3}] 33/[\ion{Si}{2}] 34~\micron\ line ratio 
within the observed range.
}
\end{figure*}

To learn more about the ionizing SEDs in Sgr~B1, we compared the line ratios 
and ionic ratios ({ionic ratios being more reliable because they} account for variations in density) 
observed in the selected sub-regions (Table~1 and Figure~2) to grids of \ion{H}{2} region models.
By plotting Ne$^{++}$/O$^{++}$ versus O$^{++}$/S$^{++}$ (ion or line ratio), 
we sample the 23--63 eV portion of the extreme ultra-violet SED
as found in both the stars that ionize \ion{H}{2}~regions and 
the stellar atmosphere models used in \ion{H}{2}~region models. 
Such comparisons inform us both about the $T_{\rm eff}$ of the \ion{H}{2}~region's exciting stars 
and about the completeness of the codes used to compute the stellar atmosphere model SEDs 
(Simpson et al. 2004; Rubin et al. 2008).
The averages of the sub-regions (Table~1) are plotted in Figure~3, 
along with the ratios from the models. 

{ 
These models include the coarse grid of \ion{H}{2} region models from S18.
Her models all had hydrogen densities $N_p = 100$~cm$^{-3}$, 
ionizing photon luminosities of 10$^{50}$~s$^{-1}$, 
GC abundances (S18), 
filling factors ranging from 0.001 to 1.0, 
\ion{H}{2} region inner radii ranging from 1 -- 10~pc, 
and ionizing stellar SEDs from Starburst99 (Leitherer et al. 1999, 2014) with ages ranging from 
$10^6$ to $10^{6.7}$ yr and augmented with X-rays.
For computational ease, the X-rays were represented by black bodies with temperatures, $T_{BB}$,
of either $10^{6.0}$ or $10^{6.5}$~K and black-body luminosities, $L_{BB}$, ranging from 
$10^{37}$ to $10^{39}$~erg~s$^{-1}$. However, 
because the line that required the X-rays for fitting was the [\ion{O}{4}] 26~\micron\ line 
with ionization potential 54.9 -- 77~eV, only the X-rays in this energy range 
were significant to the model fits. 
The integrated ionizing-luminosity from 55 --~77~eV is $1.9 \times 10^{35}$ or $6.2 \times 10^{35}$ erg~s$^{-1}$ 
for the two best-fitting models for Sgr~B1 (input $3 \times 10^{50}$~s$^{-1}$ ionizing photons, $T_{BB} = 10^{6.5}$~K, 
and $L_{BB} = 3 \times 10^{38.0}$ or $3 \times 10^{38.5}$~erg~s$^{-1}$, respectively). 
Ratios from the models augmented with X-rays are plotted with asterisks in Figure~3
(omitting all models with predicted [\ion{S}{3}] 33/[\ion{Si}{2}] 34~\micron\ line ratios $<  0.75$).

Additional models with no X-rays were computed for this paper using Cloudy 17.01 (Ferland et al. 2017). 
The models using stellar SEDs from Starburst99 have the same variation in SED age, inner radius, 
and filling factor as the models with X-rays but $N_p = 300$~cm$^{-3}$.
These models are plotted with squares in Figure 3, where it is seen that 
none of these models has line ratios that lie in the region of the observed line ratios.
We conclude that the only models that use Starburst99 SEDs that also have X-rays agree with the data.

However, the SEDs from Starburst99 are the summation of the individual stellar SEDs from a massive cluster of stars, starting with an initial mass function (IMF) and evolved along prescribed evolutionary tracks (Leitherer et al. 1999).
Because Sgr~B1 apparently does not consist of a centralized cluster but instead contains a number of individual \ion{H}{2}~regions each containing too few stars to completely sample the original IMF, 
separate modeling of each of the sub-regions can elucidate the characteristics of their ionizing stars (e.g., Simpson et al. 2004).
Consequently, a set of models (plotted as triangles) was computed using the same input density, inner radii, and filling factors and using stellar SEDs computed with the WM-BASIC code (Pauldrach et al. 2001) 
by the Binary Population and Spectral Synthesis (BPASS) project (Eldridge et al. 2017).
In Figure 3 the WM-BASIC models that have the observed values of either O$^{++}$/S$^{++}$
or [\ion{O}{3}] 88/[\ion{S}{3}] 33~\micron\ all have $T_{\rm eff}$ of 32300 or 34600~K;
for the WM-BASIC code, these $T_{\rm eff}$ correspond to supergiants with spectral types O8.5 -- O9 
and even later spectral types for dwarfs (Sternberg et al. 2003).
From the lack of O stars earlier than this, we infer ages of several Myr, 
in agreement with the ages estimated from the comparison with Starburst99 SEDs.

On the other hand, there could possibly be contributions from high-velocity shocks. 
Ho et al. (2014) observed optical forbidden lines in galactic outflows 
in the star-forming galaxy SDSS J090005.05+000446.7.
They found that their models fit the observed line ratios better if they added various amounts of shock-model line intensities 
to the line intensities predicted by their photoionization models. 
Following their procedure, we plot in Figure 3(b) the effects of adding  
shocked-gas intensities to a sample Cloudy model.
These shock models from Allen et al. (2008) are their 
`L\_n100\_b0.01' (density 100 cm$^{-3}$, $B = 0.01~\mu$G)  % mapp2 cyan
and `S\_n1000\_b0.01' (density 1000 cm$^{-3}$, $B = 0.01~\mu$G) model sets, %maps1  brown
plotted as cyan and brown lines, respectively.
They were chosen because their neutral gas densities bracket the observed ionized gas densities 
and because they came closest to producing the observed [\ion{S}{3}] 33/[\ion{Si}{2}] 34~\micron\ line ratios, compared to models with larger $B$.
Both series of shock models have nominally solar abundance, but the line ratios were adjusted 
for the GC abundances from S18, which are not very different from solar.
In Figure 3(b), the fractional amounts of shock model lines are added to the model 
overplotted with the large gray cross (a model with a WM-BASIC supergiant SED with $T_{\rm eff} = 30200$~K),
chosen only to make it easier to distinguish the effects of adding shocks from the effects of adding X-rays (other models could be used with similar results).
Here, the connected points indicate increasing shock velocity 
ranging from 100, 125, ... 350 km s$^{-1}$ (left to right) 
and the unconnected points (bottom to top) indicate 
the various fractional additions of  0.01, 0.05, 0.1, 0.3, 0.5, and 1.0 
 times the line intensities.
%Note that small amounts of added shock excitation mainly increase the [\ion{Ne}{3}] 15.6/[\ion{O}{3}] 88 \micron\ line ratio, and increasing the shock velocity mainly increases the [\ion{O}{3}] 88/[\ion{S}{3}] 33 \micron\ line ratio. 
As expected, the effects of adding shocks are similar to the effects of adding X-rays, no doubt because shocks produce very hot gas, which cools by radiating X-ray thermal bremsstrahlung (e.g., Allen et al. 2008).
}

In both panels of Figure~3, we see that the observed Ne$^{++}$/O$^{++}$ ratios 
or the [\ion{Ne}{3}] 15.6/[\ion{O}{3}] 88~\micron\ line ratios 
with few exceptions agree only with models with additional X-rays.
This includes models where the X-rays originate in fast shocks (Allen et al. 2008), 
as seen in Figure~3(b). 
Shocks are especially likely for Position~1, which includes some high-energy [Ne V] emission 
($IP = 97$ eV; Table~8 of S18), 
and Positions~7 and 8, which are locations of regions with multiple radial-velocity components 
in the radio-recombination-line measurements of M92.

For the cluster age or stellar $T_{\rm eff}$, the plotted observations 
are found mostly in the vicinity of models 
with either Starburst99 age $= 10^{6.65} - 10^{6.60}$ yr 
or $T_{\rm eff} = 32300 - 34600$~K for WM-BASIC model atmospheres (Eldridge et al. 2017).
From this, we conclude that the Starburst99 age of $\sim 4.6$ Myr for Sgr~B1 of S18 
is not an anomaly due to the choice of line ratios used in the analysis 
since we now infer a similar age with the addition of the very important O$^{++}$ ionization state. 

\section{Summary and Conclusions}

We have mapped the GC \ion{H}{2} region Sgr~B1 with FIFI-LS on {\it SOFIA} in the lines of [\ion{O}{3}] 
52 and 88~\micron.
From the extinction-corrected line ratios, we computed a generally low  
density, even in regions of high radio emission, 
with no high-density clumps indicating active or incipient massive-star formation. 

We next compared the [\ion{O}{3}] 88~\micron\ line intensities 
to intensities of [\ion{S}{3}] and [\ion{Ne}{3}] 
lines taken from archived {\it Spitzer} measurements by S18. 
These ionic ratios indicate that there is no central ionizing source but instead 
at least eight small sub-regions with higher ionization surrounded by larger expanses of 
low-ionization gas. 
The regions of higher ionization have only a small correlation with the regions of higher density. 
By comparing the line ratios and ratios with \ion{H}{2} region models computed with Cloudy 
(Ferland et al. 2017), we find that the results of S18 are confirmed --- 
the region is ionized either by late O stars with $T_{\rm eff} < 35000$~K 
or by stellar clusters with ages $\sim 4 - 5$ Myr (SEDs were calculated with Starburst99, 
Leitherer et al. 2014).
X-rays in addition to those of the stellar SEDs are required to produce the measured 
Ne$^{++}$/O$^{++}$ ratio. 
X-rays from fast shocks, such as those computed by Allen et al. (2008), also can produce 
the observed line ratios. 
We conclude that having lines from ions with $IP \sim35$~eV  
enables a significant contribution to our understanding of an \ion{H}{2} region's ionizing SED --- 
even the Ne$^{++}$ ionization stage ($IP = 41$ eV) can show the effects of local X-rays and/or shocks.

We conclude that the scattered locations of the ionizing stars and their inferred low $T_{\rm eff}$, 
in addition to the low densities 
and apparent dispersal of the ionized gas (e.g., M92), 
all indicate that any ionizing star cluster is at least a few Myr old, 
in spite of the close velocity connection of the gas to the young, star-forming region Sgr~B2.
We suggest that the stars that ionize Sgr~B1 were not formed in situ, 
but date from a previous era of star formation 
and have already orbited the Galactic nucleus back to their present location at 
Galactic longitude $\sim 0\fdg5$. 
This scenario is similar to the suggested origin of the Arches and Quintuplet Clusters. 

\acknowledgments
Based on observations made with the NASA/DLR Stratospheric Observatory for Infrared Astronomy (SOFIA). SOFIA is jointly operated by the Universities Space Research Association, Inc. (USRA), under NASA contract NNA17BF53C, and the Deutsches SOFIA Institut (DSI) under DLR contract 50 OK 0901 to the University of Stuttgart.
Financial support for this work was provided by NASA through awards 04-0113 and 05-0082 issued by USRA. 
We thank Christian Fischer and Randolf Klein for assistance with the observations 
{ and the referee for the thoughtful comments that improved the presentation of the paper}.

\vspace{5mm}
\facility{SOFIA(FIFI-LS)}

\software{Cloudy \citep{cloudy17}
}


\begin{thebibliography}{}

\bibitem[Allen et al.(2008)]{allen08}Allen, M. G., Groves, B. A., Dopita, M. A., Sutherland, R. S., \& Kewley, L. J. 2008, \apjs, 178, 20

\bibitem[An et al.(2011)]{an11}An, D.,  Ram\'irez, S. V., Sellgren, K., et al. 2011, \apj, 736, 133

\bibitem[An et al.(2017)]{an17}An, D., Sellgren, K., Boogert, A. C. A., Ram\'irez, S. V., \& Pyo, T. S. 2017, \apj, 843, 36

\bibitem[Barnes et al.(2017)]{barnes17}Barnes, A. T., Longmore, S. N., Battersby, C., et al. 2017, \mnras, 469, 2263

\bibitem[Colditz et al.(2018)]{fifi-cal}Colditz, S., Beckmann, S., Bryant, A., et al. 2018, J. Astron. Instrum., 7, 1840004

\bibitem[De Pree et al.(2015)]{depree15}De Pree, C. G., Peters, T., Mac Low, M. M., et al. 2015, \apj, 815, 123
\bibitem[Eldridge et al.(2017)]{BPASS}Eldridge, J. J., Stanway, E. R., Xiao, L., et al. 2017, \pasa, 34, 58

\bibitem[Ferland et al.(2017)]{cloudy17}{Ferland, G. J., Chatzikos, M., Guzm\'an, F., et al. 2017, \rmxaa, 53, 385}

\bibitem[Fischer et al.(2018)]{fifi-ls}Fischer, C., Beckmann, S., Bryant, A., et al. 2018, J. Astron. Instrum., 7, 1840003


\bibitem[Habibi et al.(2014)]{habibi14}{Habibi, M., Stolte, A., \& Harfst, S. 2014, \aap, 566, A6}

\bibitem[Ho et al.(2014)]{ho2014}Ho, I.-T., Kewley, L. J., Dopita, M. A., et al. 2014, \mnras, 444, 3894

\bibitem[Kruijssen et al.(2015)]{kruijssen15}Kruijssen, J. M., D., Dale, J. E., \& Longmore, S. N. 2015, \mnras, 447, 1059

%\bibitem[Kruijssen et al.(2014)]{kruijssen14}Kruijssen, J. M. D., Longmore, S. N., Elmegreen, B. G., et al. 2014, \mnras, 440, 3370


\bibitem[Lang et al.(2010)]{lang10}Lang, C. C., Goss, W. M., Cyganowski, C., \& Clubb, K. I. 2010, \apjs, 191, 275
%\bibitem[Lang et al.(2001)]{lgm01} Lang, C. C., Goss, W. M., \& Morris, M. 2001, \aj, 121, 2681
\bibitem[Lang et al.(2002)]{lgm02} Lang, C. C., Goss, W. M., \& Morris, M. 2002, \aj, 124, 2677
\bibitem[Leitherer et al.(2014)]{starburst2014}Leitherer, C., Ekstr\"om, S., Meynet, G., et al. 2014, \apjs, 212, 14
\bibitem[Leitherer et al.(1999)]{starburst1999}{ Leitherer, C., Schaerer, D., Goldader, J. D., et al. 1999, \apjs, 123, 3}

\bibitem[Longmore et al.(2013)]{longmore13}Longmore, S. N., Kruijssen, J. M. D., Bally, J., et al. 2013, \mnras, 433, L15


\bibitem[Mauerhan et al.(2010)]{mauerhan10}Mauerhan, J. C., Muno, M. P., Morris, M. R., Stolovy, S. R., \& Cotera, A. 2010, \apj, 710, 706

\bibitem[Mehringer et al.(1993)]{mehringer93}Mehringer, D. M., Palmer, P., \& Goss, W. M. 1993, \apj, 402, L69
\bibitem[Mehringer et al.(1995)]{mehringer95}Mehringer, D. M., Palmer, P., \& Goss, W. M. 1995, \apjs, 97, 497
\bibitem[Mehringer et al.(1992)]{mehringer92}Mehringer, D. M., Yusef-Zadeh, F., Palmer, P., \& Goss, W. M. 1992, \apj, 401, 168 (M92)

\bibitem[Molinari et al.(2011)]{molinari11}Molinari, S., Bally, J., Noriega-Crespo, A., et al. 2011, \apj, 735, L33
\bibitem[Molinari et al.(2016)]{molinari16}Molinari, S., Scisano, E., Elia, D., et al. 2016, \aap, 591, A149

\bibitem[Pauldrach et al.(2001)]{pauldrach01} Pauldrach, A. W. A., Hoffmann, T. L., \& Lennon, M. 2001, \aap, 375, 161

\bibitem[Rubin et al.(2008)]{rubin08} Rubin, R. H., Simpson, J. P., Colgan, S. W. J., et al. 2008, \mnras, 387, 45

\bibitem[Schneider et al.(2014)]{schneider14}Schneider, F. R. N., Izzard, R. G., de Mink, S. E., et al. 2014, \apj, 780, 117

\bibitem[Simpson(2018)]{simpson18}Simpson, J. P. 2018, \apj, 857, 59 (S18)
\bibitem[Simpson et al.(2012)]{simpson12} Simpson, J. P., Cotera, A. S., Burton, M. G., et al. 2012, \mnras, 419, 211
\bibitem[Simpson et al.(2007)]{simpson07}Simpson, J. P., Colgan, S. W. J., Cotera, A. S., et al. 2007, \apj, 670, 1115


\bibitem[Simpson et al.(2004)]{src04} Simpson, J. P., Rubin, R. H., Colgan, S. W. J., Erickson, E. F., \& Haas, M. R. 2004, \apj, 611, 338
\bibitem[Sternberg et al.(2003)]{sternberg03}{ Sternberg, A., Hoffmann, T. L., \& Pauldrach, A. W. A. 2003, \apj, 599, 1333}

\bibitem[Stolte et al.(2014)]{stolte14}Stolte, A., Hu{\ss}mann, B., Morris, M. R., et al. 2014, \apj, 789, 115
\bibitem[Storey \& Sochi(2015)]{storey15} Storey, P. J., \& Sochi, T. 2015, \mnras, 449, 2974

\bibitem[Temi et al.(2014)]{temi14}Temi, P., Marcum, P. M., Young, E., et al. 2014, \apjs, 212, 24

\bibitem[Young et al.(2012)]{young12}Young, E. T., Becklin, E. E., Marcum, P. M., et al. 2012, \apj, 749, L17

\end{thebibliography}
\end{document}